\documentclass[aps,pre,twocolumn,showpacs,showkeys,floatfix]{revtex4}
\usepackage{graphicx}% Include figure files
\usepackage{dcolumn}% Align table columns on decimal point
\usepackage{bm}% bold math

\newcommand{\diffl}[2]{\frac{d #1}{d #2}}
\newcommand{\dfrac}[2]{\displaystyle\frac{#1}{#2}}

\begin{document}

%\preprint{Manuscript ??????}

\title{Simultaneous compression and opacity data from time-series radiography \\
   with a Lagrangian marker}

\date{April 5, 2018; revisions to October 15, 2020
%  -- IM\#1009466}
  -- LLNL-JRNL-805296}

\author{Damian C. Swift}
\email{dswift@llnl.gov}
%\homepage{http://public.lanl.gov/dswift}
\affiliation{%
   Lawrence Livermore National Laboratory,
   7000 East Avenue, Livermore, California 94550, USA
}
\author{Andrea L. Kritcher}
\affiliation{%
   Lawrence Livermore National Laboratory,
   7000 East Avenue, Livermore, California 94550, USA
}
\author{James A. Hawreliak\footnote{%
Current affiliation:
Washington State University
}}
\affiliation{%
   Lawrence Livermore National Laboratory,
   7000 East Avenue, Livermore, California 94550, USA
}
\author{James Gaffney}
\affiliation{%
   Lawrence Livermore National Laboratory,
   7000 East Avenue, Livermore, California 94550, USA
}
\author{Amy Lazicki}
\affiliation{%
   Lawrence Livermore National Laboratory,
   7000 East Avenue, Livermore, California 94550, USA
}
\author{Andrew MacPhee}
\affiliation{%
   Lawrence Livermore National Laboratory,
   7000 East Avenue, Livermore, California 94550, USA
}
\author{Benjamin Bachmann}
\affiliation{%
   Lawrence Livermore National Laboratory,
   7000 East Avenue, Livermore, California 94550, USA
}
\author{Stephen D. Rothman}
\affiliation{%
   Atomic Weapons Establishment,
   Aldermaston, Berkshire, RG7~4PR, UK
}
\author{Dominik Kraus\footnote{%
Current affiliation: Helmholtz Zentrum, Dresden
}}
\affiliation{%
   University of California -- Berkeley,
   California 94720, USA
}
\author{Tilo D\"oppner}
\affiliation{%
   Lawrence Livermore National Laboratory,
   7000 East Avenue, Livermore, California 94550, USA
}
\author{Joseph Nilsen}
\affiliation{%
   Lawrence Livermore National Laboratory,
   7000 East Avenue, Livermore, California 94550, USA
}
\author{Heather D. Whitley}
\affiliation{%
   Lawrence Livermore National Laboratory,
   7000 East Avenue, Livermore, California 94550, USA
}
\author{Gilbert W. Collins\footnote{%
Current affiliation:
University of Rochester
}}
\affiliation{%
   Lawrence Livermore National Laboratory,
   7000 East Avenue, Livermore, California 94550, USA
}
\author{Siegfried Glenzer\footnote{%
Current affiliation:
SLAC National Accelerator Laboratory
}}
\affiliation{%
   Lawrence Livermore National Laboratory,
   7000 East Avenue, Livermore, California 94550, USA
}
\author{Roger W. Falcone}
\affiliation{%
   University of California -- Berkeley,
   California 94720, USA
}

\begin{abstract}
Time-resolved radiography can be used to obtain absolute shock
Hugoniot states by simultaneously measuring at least two mechanical
parameters of the shock, and this technique is particularly suitable
for one-dimensional converging shocks where a single experiment probes 
a range of pressures as the converging shock strengthens.
However, at sufficiently high pressures, the shocked material becomes
hot enough that the x-ray opacity falls significantly.
If the system includes a Lagrangian marker, such that the mass within the
marker is known, this additional information can be used to constrain the
opacity as well as the Hugoniot state.
In the limit that the opacity changes only on shock heating, and not 
significantly on subsequent isentropic compression, the opacity of 
shocked material can be determined uniquely.
More generally, it is necessary to assume the form of the variation of
opacity with isentropic compression, or to introduce multiple marker layers.
Alternatively, assuming either the equation of state or the opacity,
the presence of a marker layer in such experiments enables the non-assumed
property to be deduced more accurately than from the radiographic 
density reconstruction alone.
An example analysis is shown for measurements of a converging shock wave
in
polystyrene,
at the National Ignition Facility.
\end{abstract}

% 05.70.Ce Thermodynamic functions and equations of state
% 07.35.+k High-pressure apparatus; shock tubes; diamond anvil cells
%% fluid dyn -> compressible flows, shock and detonation phenom: 47.40.-x
% 47.40.Nm Shock wave interactions and shock effects
%% 62.50.+p High-pressure and shock wave effects in solids and liquids
% 51.30.+i thermodynamic properties, equations of state
% 52.38.Mf Laser ablation
% 64.30.-t EOS of specific substances
% 64.30.Kj EOS of nonmetals
%\keywords{Suggested keywords}%Use showkeys class option for keyword display

\maketitle

\section{Introduction}
States of matter at elevated pressure and temperature are often generated
in shock wave experiments \cite{shock}, 
where the high-pressure matter is confined inertially
(i.e. by the finite time required for the compressed components
to disassemble) and so the pressures achieved are not limited by the 
strength of surrounding components as is the case with static presses
\cite{dac}.
Material strength ranges up to $\sim$100\,GPa, and the use of shock
and other dynamic loading
experiments is ubiquitous for high energy density studies of warm dense
matter where pressures in excess of 1\,TPa are of interest.

Large pulsed lasers such as the National Ignition Facility (NIF) \cite{NIF}
and Laboratory for Laser Energetics ({\sc omega} laser, University of Rochester)
\cite{OMEGA}
can be used to induce pressures in excess of 10\,TPa, which are
of interest for studies of massive exoplanets, brown dwarfs, and stars
\cite{exoplanets,Basri2006,Swift2012,Kirkpatrick2005,Arndt1998},
as well as engineering problems such as inertial confinement fusion (ICF)
\cite{icf}.
We have previously reported the use of radiography of a spherically-converging
shock to deduce a range of states along the shock Hugoniot of a material
\cite{Kritcher2014,Doeppner2018,Swift2018}, up to $\sim$12\,TPa in polystyrene,
and thus to constrain the equation of state (EOS).
The x-ray source was a foil, laser-heated to a plasma emitting strongly in
the kilovolt regime.
In the results reported previously, the shock temperature was low enough
that the $K$-shell of the carbon atoms in the sample was not ionized 
significantly, which would have reduced the x-ray opacity and complicated
the determination of mass density from the radiograph.
However, at higher shock temperatures, 
the $K$-shell becomes significantly ionized.
This ionization could occur at lower shock pressures if the initial mass density of the
sample is lower, so this effect is relevant beyond the relatively high
pressures considered here.

In the work reported here, we consider the use of a Lagrangian marker layer
to constrain the mass enclosed, and hence enable the opacity to be deduced
along with the mass density.
The results reported here extend our previously-reported measurements
\cite{Doeppner2018,Swift2018} to higher pressures, from experiments
with a higher drive energy.
This article describes the analysis procedure and results referred to in
our recent article discussing the measurements and their relevance to
white dwarf stars \cite{Kritcher2020}.
%Although measurements have been made on polystyrene, which extend
%the previously-reported results to higher pressures, we performed an experiment
%on deutero-polyethylene 
%(which we will refer to in defiance of chemical rigor as CD$_2$) 
%which was more recent and the results from which are
%more suitable for demonstrating the simultaneous inference of mass density
%(and hence an absolute determiation of the shock Hugoniot) and opacity.

\section{Experimental configuration}
\begin{figure}
\begin{center}
\includegraphics[scale=0.30]{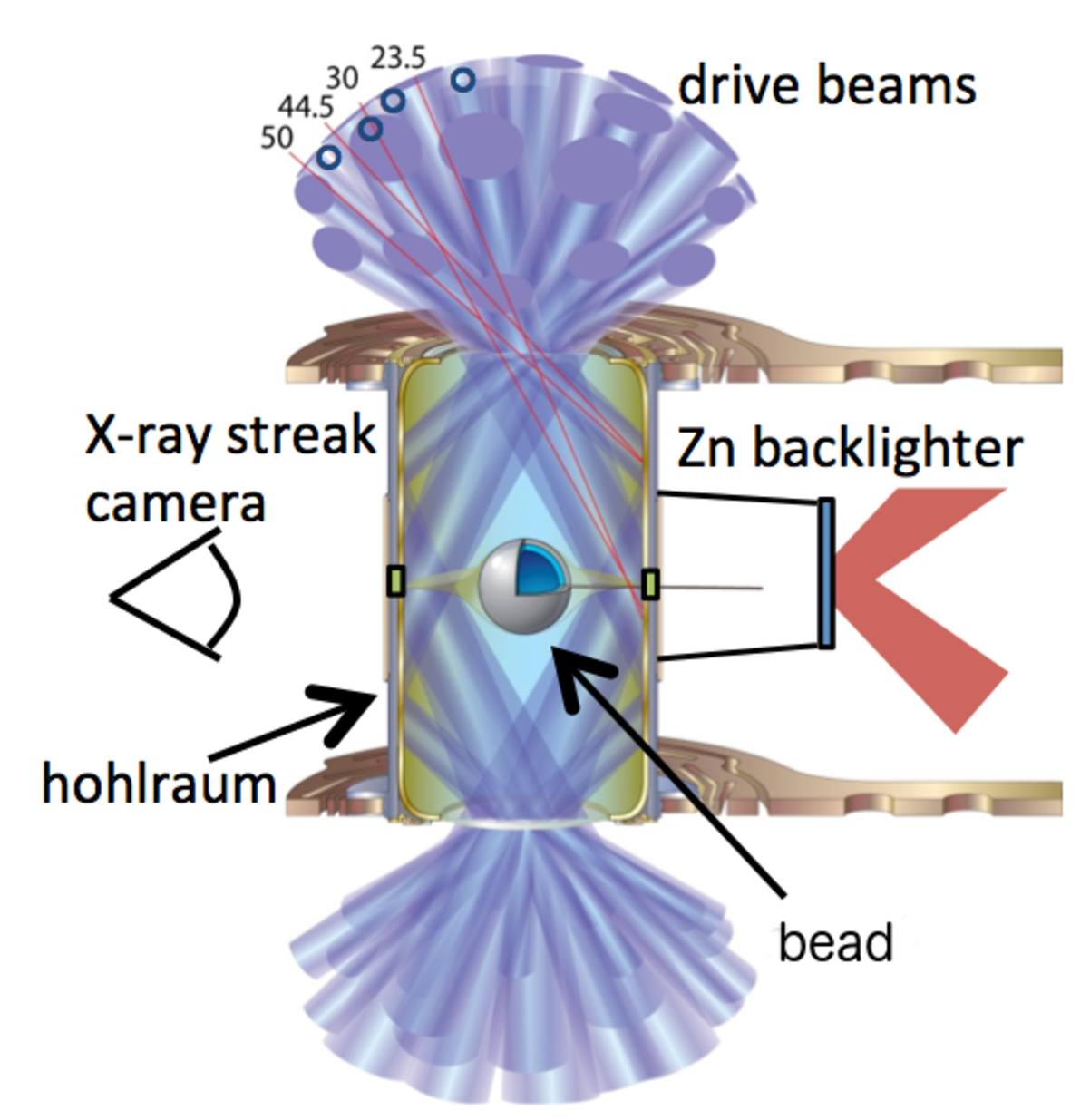}
\includegraphics[scale=0.20]{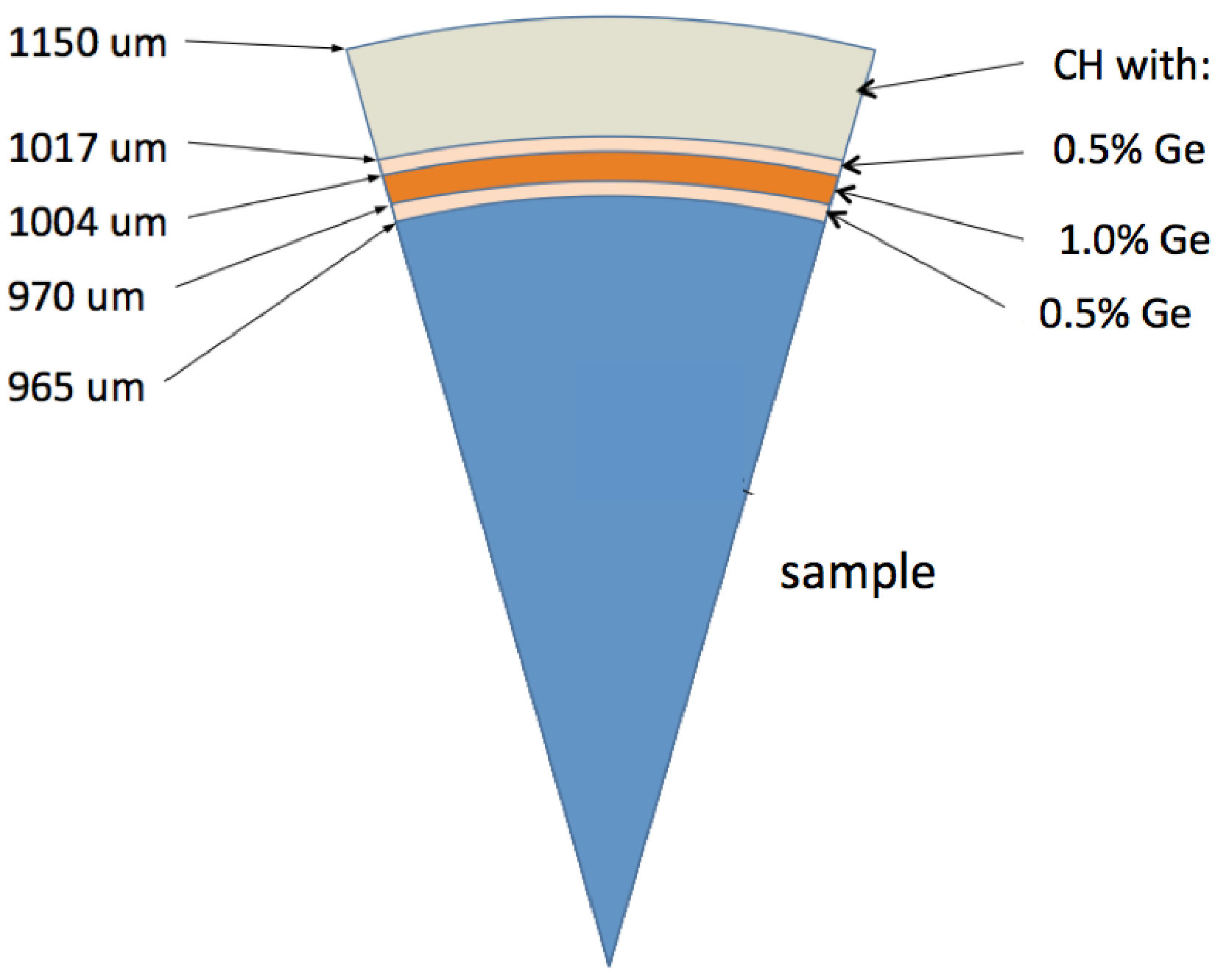}
\end{center}
\caption{Schematic of hohlraum-driven converging-shock experiment.
   Wedge diagram shows sequence of shells comprising spherical target bead.}
\label{fig:exptschem}
\end{figure}

The experimental configuration was as described previously
\cite{Kritcher2014,Swift2018,Doeppner2018},
and is summarized here for convenience.
The sample material was in the form of a solid sphere, surrounded by a
shell of glow-discharge polymer (GDP) to act as an ablator, together
referred to as the bead.
The bead was mounted within a Au hohlraum \cite{hohlraum}.
With the exception of some beams used to generate x-rays for radiography,
the remainder of
the 192 beams of the NIF laser were used to heat the hohlraum and thus 
drive the bead.
The hohlraum was filled
with He to reduce the rate that ablated Au could permeate the hohlraum
and impede the propagation of the laser beams.
The resulting soft x-ray field within the hohlraum ablated the GDP,
driving a shock into the bead.
The overall configuration and laser pulses were based on ICF designs,
to take advantage of synergies in fabrication and also the large development
effort performed to give uniform drive conditions over the surface of the
bead \cite{icf,Kritcher2014}.
(Fig.~\ref{fig:exptschem}.)

We consider data from two experiments, N130103-1 and N130701-1.
In both cases, the sample was poly(alpha-methyl styrene) (PaMS),
coated with a standard ICF ablator comprising GDP with a radially-varying
concentration of Ge (Fig.~\ref{fig:exptschem}) designed to absorb
$M$-band radiation from the hohlraum.
The hohlraum itself was Au, 30\,$\mu$m thick, 5.75\,mm diameter and 9.42\,mm high.
For commonality with the ICF campaign, the target in N130103-1 was cooled to 24\,K,
with a gas fill of 0.96\,mg/cm$^3$ He.
When cooled, the initial mass density of the PaMS was calculated to be 1.13\,g/cm$^3$.
Following changes to the allowed NIF experimental configurations, the target in N130701-1
was fired at ambient temperature, with a gas fill of 0.03\,mg/cm$^3$ He.
The temporal shape of the laser pulses heating the hohlraum was based on ICF studies
on the symmetric implosion of hollow capsules \cite{icf,Hopkins2015},
which were designed to induce a pressure history in the ablator comprising 
a series of shocks to successively higher pressures.
The pulse shape in N130103-1 induced four shocks, compared with two
in N130701-1, again employing the most appropriate ICF configuration 
available at the time of each shot.
With our solid sample, the shocks coalesced just within the sample to form a single,
strong shock, so the precise pressure history induced in the ablator did not
matter.
The temperature history of soft x-rays in the hohlraum was calculated by
radiation hydrodynamics using the {\sc hydra} program \cite{hydra}, 
and measured by the {\sc dante} filtered diode system
\cite{Dewald2004}.
The peak temperature was around 275\,eV.

\begin{figure}
\begin{center}\includegraphics[scale=0.65]{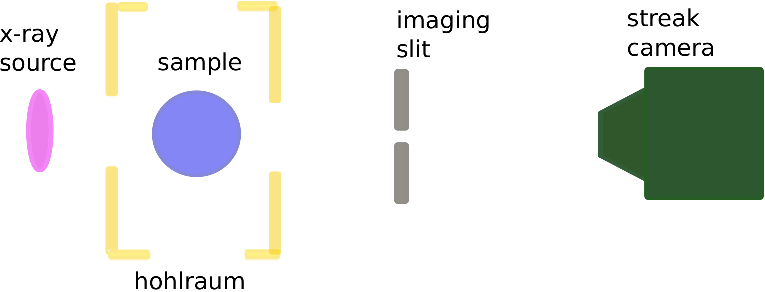}\end{center}
\caption{Radiograpic configuration (not to scale).}
\label{fig:rgconfig}
\end{figure}

The shock wave induced by ablation of the GDP strengthened as it propagated
toward the center of the sample.
X-ray radiography was used to measure the variation of attenuation
across the diameter of the bead, from which the shock trajectory,
mass distribution and opacity in the sample could be deduced as described
below.
The x-ray source was a Zn foil, driven by several laser beams 
to produce a plasma
that emitted strong He-like radiation (i.e. from atoms stripped of all but
two electrons).
Eight beams were used in N130103-1, increased to sixteen in N130701-1
to increase the x-ray signal.
Slits were cut in the hohlraum wall to enable transmission of the x-rays
through the sample; the slits were filled with diamond wedges to impede
their closure by ablated Au.
The transmitted x-rays were imaged through a slit in a Ta foil onto 
an x-ray streak camera (Fig.~\ref{fig:rgconfig}).

The streak radiograph was used to reconstruct the radial distribution
of mass density, as a function of time.
As described previously \cite{Swift2018}, the presence of undisturbed material
ahead of the shock provided a strong constraint on the inference of
the change in attenuation across the shock front.
In order to take advantage of this constraint, the analysis was performed by
adjusting a parameterized representation of the distribution of mass density
until the corresponding simulated radiograph matched the measured radiograph.
In the previous study, the change in attenuation was related directly to
a change in mass density.
At the higher pressures studied here, the change in attenuation was
a combination of the change in mass density (increasing the attenuation)
and in opacity (decreasing the attenuation).

\begin{figure}
\begin{center}\includegraphics[scale=0.47]{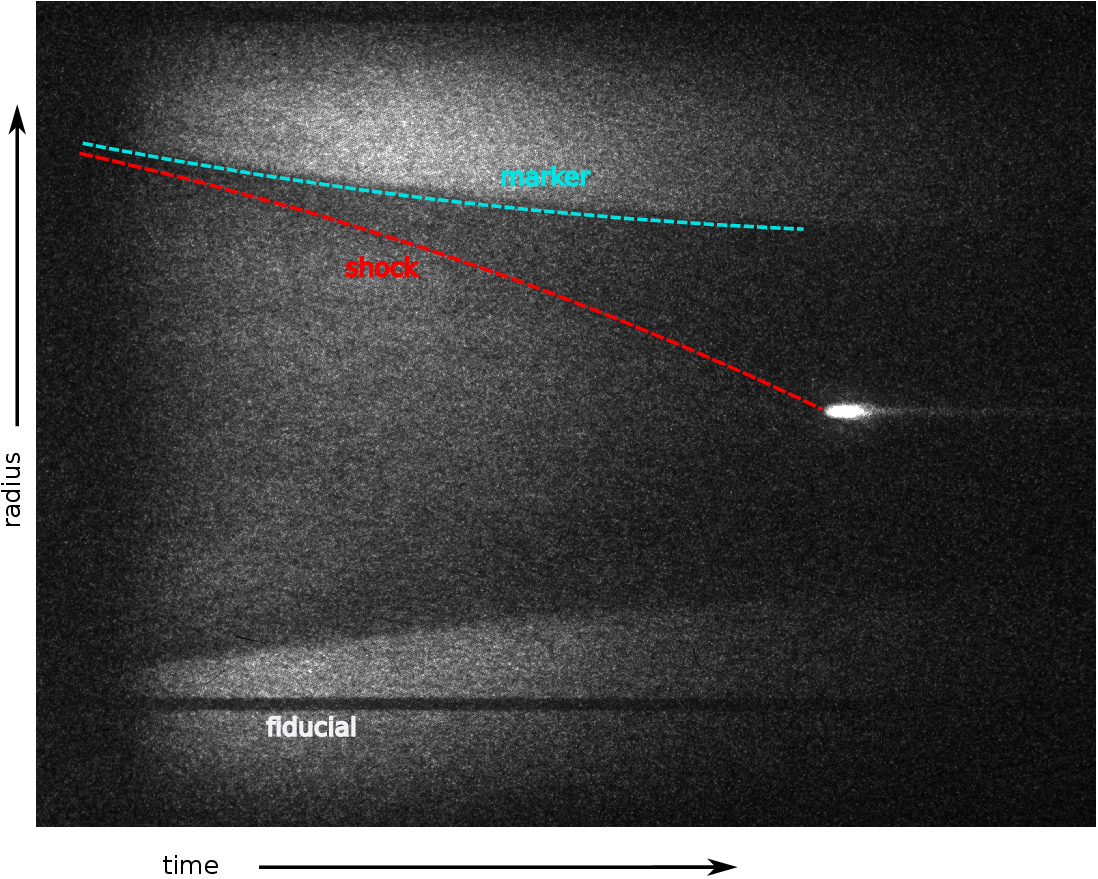}\end{center}
\caption{X-ray streak radiograph, NIF shot N130103-1 (cryogenic).}
\label{fig:rg130103}
\end{figure}

\begin{figure}
\begin{center}\includegraphics[scale=0.48]{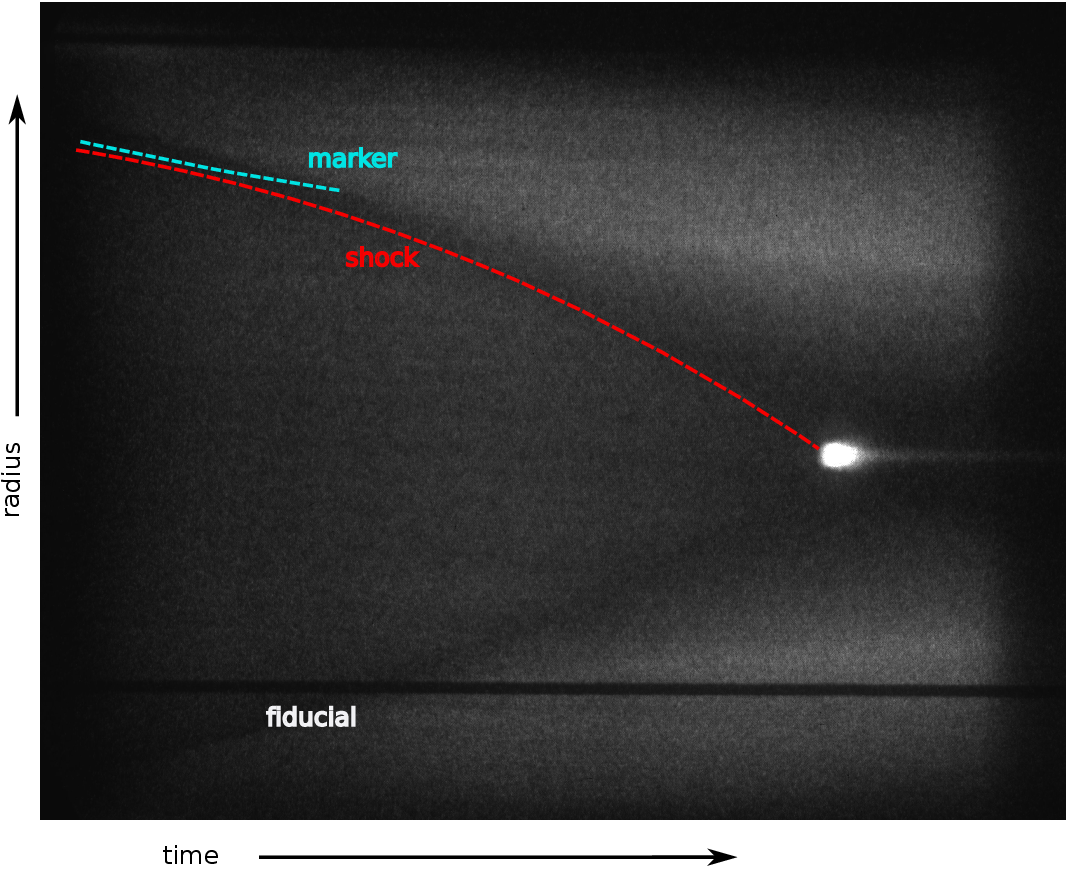}\end{center}
\caption{X-ray streak radiograph, NIF shot N130701-1 (ambient temperature).}
\label{fig:rg130701}
\end{figure}

Crucially for this experiment, the GDP ablator included a region doped
with Ge, which was visible on the radiograph from shot N130103-1 (Fig.~\ref{fig:rg130103}).
As the mass inside the doped region is known, it provides an additional
constraint that can be used to infer the opacity.
In shot N130701-1, the lower hohlraum gas fill resulted in a higher intensity
of Au $M$-band radiation, which was deposited in the doped region causing it to
expand and lose contrast during the period covered by the radiograph (Fig.~\ref{fig:rg130701})\footnote{%
In a later experiment on a different plastic, 
a thicker dopant layer was used to preserve its radiographic visibility 
for longer.
}.
As a result, the radiograph from shot N130103-1 was used to deduce the shock Hugoniot for
cryogenic PaMS and its opacity as discussed below. The results were checked for
consistency with the radiograph from shot N130701-1 and with hydrocode simulations.

Another interesting feature visible in the radiographs from experiments with
high drive energies as here is a bright flash as the shock reached the
center of the bead. Hydrodynamic heating was great enough for a region
of the sample to radiate strongly enough in the kilovolt band to be
detected by the streak camera.
Radiation hydrodynamics simulations predicted x-ray intensities consistent
with the radiographs \cite{Nilsen2018}, and the compact nature of the
emitting region indicated the high degree of symmetry of the shock
\cite{Bachmann2018}.
(Figs~\ref{fig:rg130103} and \ref{fig:rg130701}.)

\section{Idealized simulations of the converging shock}
As background for the analysis method developed, it is instructive to 
consider slightly idealized simulations of the converging shock in the sample.
The simulations shown are one-dimensional Lagrangian in spherical geometry
\cite{Benson1992},
for a shock propagating into a polystyrene sphere 2\,mm in diameter,
driven with a constant pressure of 8\,TPa,
which is representative of the ablation pressure in the NIF experiments.
Polystyrene was represented by {\sc sesame} EOS 7592 \cite{ses7592}.
We consider the motion of Lagrangian tracers positioned initially at intervals
of 100\,$\mu$m through the sample.

\begin{figure}
\begin{center}\includegraphics[scale=0.68]{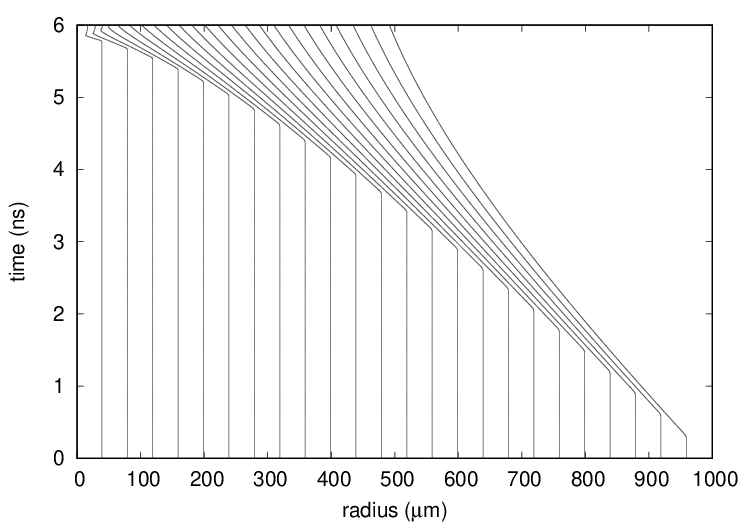}\end{center}
\caption{Motion of Lagrangian tracers in idealized simulation of 
   spherically-converging shock.}
\label{fig:tracerrt}
\end{figure}

\begin{figure}
\begin{center}\includegraphics[scale=0.68]{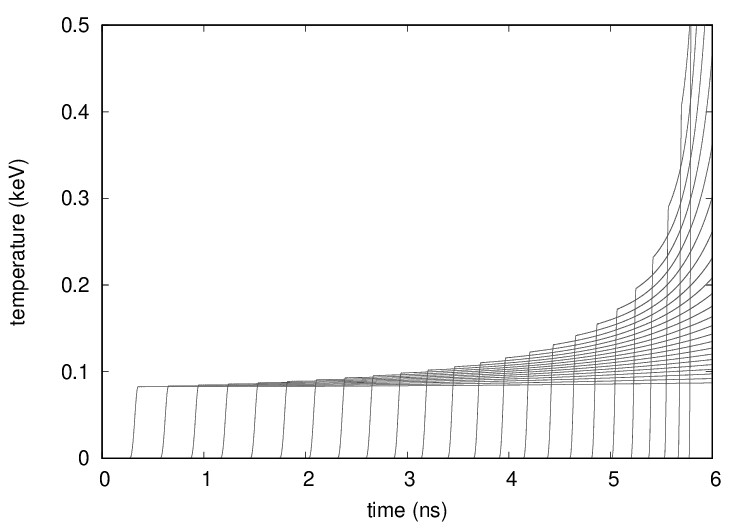}\end{center}
\caption{Temperature history of Lagrangian tracers in idealized simulation of 
   spherically-converging shock.}
\label{fig:tracertt}
\end{figure}

\begin{figure}
\begin{center}\includegraphics[scale=0.80]{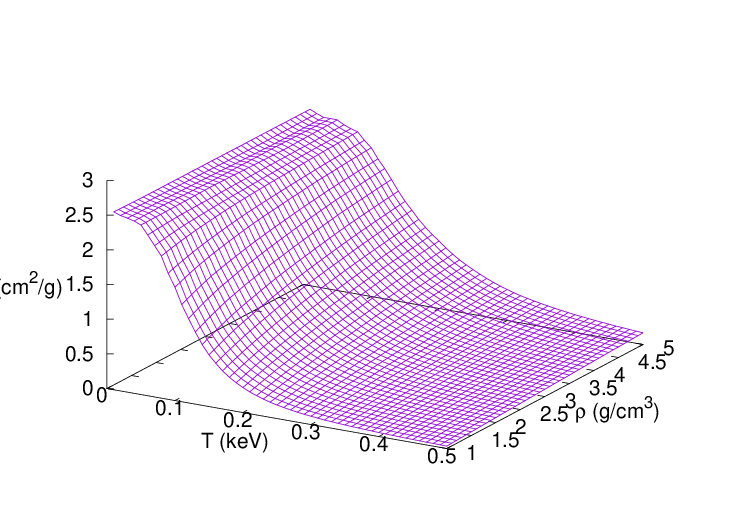}\end{center}
\caption{Opacity for polystyrene calculated using a model
   accounting for detailed configurations of excited electrons
   \cite{atomic}.}
\label{fig:thopacity}
\end{figure}

As the shock passes, each tracer is accelerated toward the center,
and the shock visibly accelerates as it nears the center
(Fig.~\ref{fig:tracerrt}).
The temperature history experienced by each tracer is a jump as the
shock passes, followed by a more gradual increase on isentropic compression
(Fig.~\ref{fig:tracertt}).
The mass density is in the range of several g/cm$^3$, and the temperature
from several tens to several hundred electron-volts.
Under these conditions, widely-used atomic models \cite{imp,opal,atomic} predict that the
opacity should vary strongly with temperature, dropping by an order of
magnitude as the shock pressure rises
(Fig.~\ref{fig:thopacity}).

% from ~/physics/shock/convergent/shock/parylene/lagc1d

% from ~/physics/radtpt/opacity/llnl

\begin{figure}
\begin{center}\includegraphics[scale=0.68]{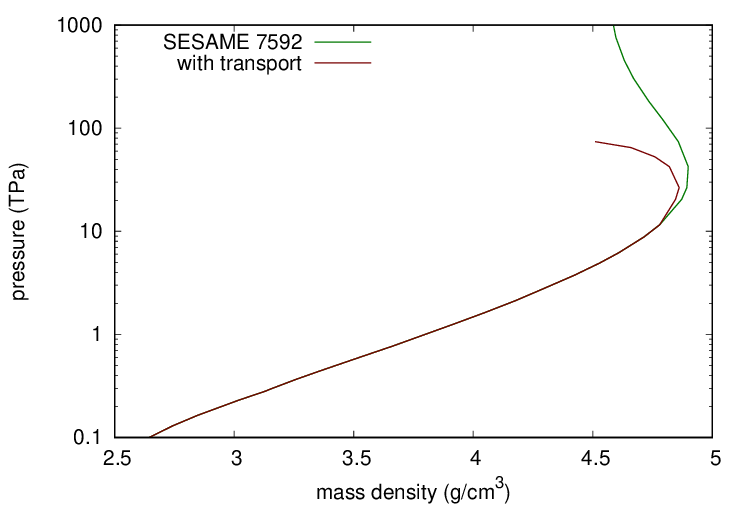}\end{center}
\caption{Sensitivity of inferred Hugoniot to heat transport ahead of the
   shock.}
\label{fig:tptsens}
\end{figure}

When the shock temperature is sufficiently high, enough energy may be
transported past the shock to heat the material ahead that the state induced
by the shock may be significantly different than the principal Hugoniot.
Using Lagrangian radiation hydrodynamics simulations with a variety of
EOS and opacity models,
we assessed the deviation in shock speed and mass density
from thermal transport.
We also assessed the change in opacity ahead of the shock, which could
affect the inferred location of the shock in the radiograph.
(Fig.~\ref{fig:tptsens}.)

% ~/physics/mats/polystyrene/eos/preheat.eps

At the temperatures accessed in these converging-shock experiments,
high enough to affect the x-ray opacity,
for all but the lowest-$Z$ elements, the thermal energy of the electrons
dominates over that of the ions, simply because there are more electrons
in the system.
Almost all wide-range EOS models have been constructed using the Thomas-Fermi
(TF) approximation of a uniform electron gas \cite{tf},
and so these EOS featured heavily in simulations used for sensitivity studies
and for comparisons with experimental data,
taken from the {\sc sesame} and {\sc leos} libraries \cite{sesame,leos}.
A key question is whether the TF approximation, or variants, is adequate,
or whether higher-order models treating electronic shell structure
are necessary.
The simplest form of shell structure treatment is the average atom model
\cite{Liberman1979,Wilson2006}, which predicts pronounced features
on the shock Hugoniot as successive electron shells are abruptly ionized;
these predictions have been questioned as potentially exaggerating these
ionization features in comparison with treatments including a more realistic 
distribution of atoms.
Electron shell effects are important when constructing models of the opacity,
which depends more sensitive than the EOS on accounting for
the distribution of occupied and vacant electron states.
As our analysis needs to use a model for off-Hugoniot variation of the opacity,
we have used a range of specific opacity models, based on calculations from
the {\sc imp} \cite{imp}, {\sc opal} \cite{opal}, and {\sc atomic} \cite{atomic}
computer programs.

Radiation hydrodynamics simulations were used in this work to guide
physical insight, design experiments, and test assumptions and analysis methods.
These simulations typically bring together the state of the art in 
EOS, opacity, and numerical methods for continuum mechanics and radiation
transport.
No one person is fully aware of every aspect of all the models and methods.
A notable benefit of exercising our simulation abilities on a new
experimental platform like this has been to identify unexpected problems and
limitations of models.
For example, we discovered that one set of opacities tabulated for low-$Z$ materials
was inaccurate for kilovolt photons
because of an incorrect extrapolation at energies far above the
binding energy of the $K$-shell.
As well as being a useful finding in its own right, it highlights the
value of designing experiments that can be interpreted as directly as possible
from the measurements themselves, without being dominated by input from models or simulations.
Even when measurements can be interpreted in this way, simulations may enter
in a subtler way by informing the choice of initial conditions for 
iterative optimization of model parameters against the experimental data.
Finally, the interplay between multiple physics models and implementations 
complicates the assessment of unexpected results, lengthening the process of
turning experimental measurements into robust conclusions.

\section{Determination of absolute Hugoniot and opacity data from radiography}
As discussed previously \cite{Swift2018},
the reconstructed radius-time distribution of mass density $\rho(r,t)$ gives 
an absolute measurement of the shock Hugoniot over a range of pressures,
from the position of the shock $r_s(t)$ and hence its speed $u_s(t)$,
and the mass density immediately behind the shock,
$\rho_s(t)=\rho(r_s(t),t)$.
Simultaneous knowledge of $u_s$ and $\rho_s$ gives the complete mechanical
state behind the shock by solving the Rankine-Hugoniot relations \cite{shock}
representing the conservation of mass, momentum, and energy across the shock,
\begin{eqnarray}
u_s^2 & = & v_0^2\dfrac{p-p_0}{v_0-v} \\
u_p & = & \sqrt{\left[(p-p_0)(v_0-v)\right]} \\
e & = & e_0 + \frac 12 (p+p_0)(v_0-v)
\end{eqnarray}
where $v$ is the reciprocal of the mass density $\rho$,
$e$ is the specific internal energy, $p$ the pressure,\footnote{%
For materials in which material strength is significant, $p$ is the
normal stress rather than the mean pressure.
}
and subscript `0' denotes material ahead of the shock (with $u_p=0$).
The state ahead of the shock is known, leaving five quantities to be
determined ($v, p, e, u_p, u_s$).
If any two of these quantities are measured, the Rankine-Hugoniot equations
determine the rest.
In particular,
\begin{equation}
p = p_0 + \dfrac{u_s^2}{v_0^2}\left(v_0-v\right).
\end{equation}
Thus the mechanical state on the Hugoniot can be deduced directly from
the distribution of mass density, without reference to any other material
used as a standard, as is the case in some other experimental configurations,
and so the measurement is absolute.

Given distributions of mass density $\rho(\vec r,t)$ and 
opacity $\sigma(\vec r,t)$ in the object, 
the signal along any path from the source $\vec r_s$ to the detector $\vec r_d$
at any instant of time is given by the integral of attenuation $\mu=\rho\sigma$
through the object,
\begin{equation}
I(\vec r_d,t)=I(\vec r_s)\exp\left[-\int_{\vec r_s}^{\vec r_d}
   \rho(\vec r,t)\sigma(\vec r,t)\,d\vec r\right]
\end{equation}
neglecting scattering.
If the opacity and density vary independently without further constraints,
their variations cannot be separated.
However, if the unknown opacity change occurs only along one locus,
such as the shock front, and if the total visible mass is known, 
the variations can be separated from time-series data because, at each instant
of time, the attenuation at the shock front and the difference between the
total mass and the apparent mass provide two measurements from which 
the two unknowns can be deduced.
The `visible mass' may be defined with respect to a feature in the
object that can be observed radiographically, to act as a 
Lagrangian marker.
Suitable features include the interface between materials 
or a thin layer of material of different opacity.

This procedure works if there is only one unknown change in opacity between 
successive radiographic frames.
For a converging shock, the compression changes behind the shock,
with accompanying changes in temperature, which may lead to a change in
the opacity.
However, as discussed above, temperature change is dominated by the shock heating
and this usually dominates the change
in opacity, i.e. $\sigma(\chi,t)\simeq\sigma(\rho;\rho_s)$.
Changes caused by subsequent adiabatic heating can be ignored or accounted for
with models of the opacity or opacity change.

Consider the radiograph as a map of the transmission through the object,
$T(x,t)$, where $x$ is the space direction in the streak record.
Neglecting scatter and the finite spatial resolution of the imaging system,
the transmission at any point $(x,t)$ in the image is found from the
integral through the object of the product of $\rho$ and the opacity $\sigma$,
\begin{equation}
T(x,t)=\exp\left[-\int \rho(r,t)\sigma(r,t)f(r,x)\,dr\right]
\end{equation}
where $f(r,x)$ represents the geometrical mapping between radius within the
object and position across the radiograph.
Analogously to profile-matching of the distribution of mass density
to match the radiograph when $\sigma$ is constant, for varying $\sigma$ 
we can instead use profile-matching to deduce the distribution
of attenuation $\mu(r,t)\equiv\sigma(r,t)\rho(r,t)$.
Although for the present experiment we are interested in spherical objects,
the same approach can be applied to planar or cylindrical systems, so
we will consider the generalized problem.

Suppose that at a given time $t_2$ in the streak radiograph, analysis of
earlier data has given the mass density and opacity,
$\{\rho,\sigma\}(r,t'<t_1)$.
We define a Lagrangian ordinate $\chi$ with respect to a reference feature
in the object (a marker layer or an edge) such that
\begin{equation}
\diffl \chi r\propto r^\alpha \rho
\end{equation}
where $\alpha$ is 0, 1, or 2 for planar, cylindrical, or spherical geometry.
The mass enclosed by the feature is $M_T$, and
the reference feature follows a trajectory $r_r(t)$.
Given the instantaneous position of the shock $r_s(t)$, the mass of material outside the shock
\begin{equation}
M_s(t)\equiv\int_{r_s(t)}^{r_r(t)} \chi(r',t)\,dr'.
\end{equation}
If the loading history experienced by each element of the sample is a shock
followed by approximately isentropic loading or unloading, the opacity
change is dominated by the amount of shock heating, 
\begin{equation}
\sigma(\chi,t>t_s(\chi))=\sigma(\rho;\rho_s).
\end{equation}
Considering the motion of the shock into initially undisturbed material
between $t_1$ and $t_2=t_1+\delta t$, we want to deduce $\rho$ for the
newly-shocked material.
The first step is to analyze the attenuation $\mu(r,t_2)$ to find the
inner radius of material shocked up to $t_1$,
\begin{equation}
r_1(t_2)\quad:\quad M_1=\int_{r_1}^{r_r}\frac{\mu(r,t_2)}{\sigma(\chi(r'))}\,dr'.
\end{equation}
If $\sigma$ is to vary with isentropic compression, the variation is 
included in the calculation of $r_1(t_2)$, and the solution may become
iterative.
From the radius of the shock $r_2$ at $t_2$, the mass of unshocked material
$M_0$, the mass density of the newly-shocked material is
\begin{equation}
\rho_s(t_2)=\frac{M_T-[M_0(t_2)+M_1(t_2)]}{V(r_1,r_2;t_2)}
\end{equation}
where $V(r_1,r_2)$ is the volume enclosed between $r_1$ and $r_2$,
which is proportional to $r_2^\alpha-r_1^\alpha$.
Thus the opacity of the newly-shocked material
\begin{equation}
\sigma(\rho_s)=\mu(r_2,t_2)/\rho_s.
\end{equation}
This Lagrangian analysis also gives $r(\chi,t)$ from $\chi(r,t)$, and hence
$u_p(\chi,t)$ and $u_p(r,t)$.

The equations above can be rewritten in differential form, but spatial
integration from the marker layer is essential to finding the solution,
so the problem has a fundamentally integro-differential character.
Because the solution involves integration between the shock and the marker,
error accumulates along the shock as the solution progresses, as with
Abel inversion \cite{Abel1826}.

To summarize, the simultaneous reconstruction of mass density and opacity
proceeds by performing profile-matching on the radiograph to determine
the attenuation $\mu(r,t)$, identifying the trajectory of the shock and the
marker, and applying the integral equations incrementally with time
to reconstruct the motion of each element of material within the
marker $r(\chi,t)$ and hence determine $\rho_s(t)$ and $\sigma_s(t)$.

For the previous analysis assuming constant opacity \cite{Swift2018},
several different methods were used to represent the radial variation of
mass density through the shocked region, including a variety
of functional forms and also tabulations with the ordinates typically 
distributed at uniform fractions of the separation between the
shock and the marker.
For the case of varying opacity, particularly when correcting for
opacity variation with isentropic compression behind the shock,
it was most efficient to construct
the variation along lines of constant $\chi$ in the shocked region.

If the opacity changes only as the shock passes, it can be determined from
the variation in the apparent mass enclosed by the marker.
\begin{equation}
\sigma_0\diffl Mt=4\pi r_s^2\rho_0 u_s(\sigma_s-\sigma_0)
\end{equation}
therefore
\begin{equation}
\sigma_s=\sigma_0\left(1+\frac{dM/dt}{4\pi r_s^2\rho_0 u_s}\right).
\end{equation}

\section{Analysis of experimental data for poly(alpha methyl styrene) (PaMS)}
In shot N130103-1, the doped layer in the ablator was visible throughout
the duration of the radiograph, so the analysis described above could be
performed to deduce the Hugoniot and opacity simultaneously.
In addition, simpler analyses were performed to deduce the Hugoniot
assuming a model for opacity,
and also to deduce the opacity assuming a model for the EOS.

The streak radiograph was analyzed to deduce
the radius-time distribution of mass density,
represented using smooth functions
as described previously \cite{Swift2018}.
The analysis was performed in several different ways to gain confidence
that the result did not depend on the choice of function.
Alternative functions were used, and the analysis was performed over the
full range of the streak record and also over shorter intervals of time.

The locus of the shock was represented by the function
\begin{equation}
r_s(t)=\alpha(t_c-t)^\beta
\end{equation}
where $\alpha$, $\beta$, and $t_c$ were fitting parameters.
The locus of the marker layer was represented by the function
\begin{equation}
r_m(t)=r_{\mbox{min}}+\alpha e^{-\beta t}
\end{equation}
where $\alpha$, $\beta$, and $r_{\mbox{min}}$ were fitting parameters.
These functions were able to capture these loci over the full range of
the record.
For shorter intervals, the same functions or polynomials were used.

The mass density in the shocked region was represented by interpolation
between functions defined along the shock locus $\rho_s(t)$ and marker layer $\rho_m(t)$,
with non-linear variation behind the shock represented either by functions defined
along intermediate loci $\rho_i(t)$ or by analytically integrable functions
\begin{equation}
f(r,t)=\alpha(t) r^{2n}\exp\left[-r^2/2\sigma(t)^2\right],
\end{equation}
The $\rho(t)$ functions and the time-dependent parameters $\alpha(t)$ and $\sigma(t)$
were represented using low-order polynomials or tabulations.

Given a reconstruction of the mass density, the Hugoniot was deduced from $\rho_s(t)$
and the shock speed, $-dr_s(t)/dt$.
The goodness-of-fit of the simulated radiograph to the data was used to assign a probability
to the model.
By perturbing the fitting parameters about the best fit, Hugoniots were deduced with 
corresponding probability.
The Hugoniot loci were accumulated as probability amplitudes.
The nominal best-fitting Hugoniot was taken to be the locus of maximum likelihood,
very similar to the Hugoniot from the best-fitting parameters,
and 1$\sigma$ uncertainties were taken as contours from the probability distribution.
These are the statistical fitting uncertainties. Systematic uncertainties from the
uncertainty in instantaneous sweep rate of the streak camera and magnification
affect the location of the Hugoniot, but affect its shape to a much smaller degree. 

The uncertainty in opacity included an analogous, statistical contribution from 
fitting the radiograph.
Since the opacity was deduced simultaneously with the reconstruction of mass density,
the value along the locus of the shock can be associated with a statistically-exact
Hugoniot state.
The uncertainty in actual Hugoniot state associated an individual opacity 
has the same statistical and systematic uncertainties as the Hugoniot state itself,
but these uncertainties are correlated exactly with the Hugoniot uncertainty.
For this reason, we show only the statistical uncertainty in opacity.

\subsection{Simultaneous Hugoniot and opacity analysis}
The Hugoniot and opacity were deduced simultaneously, first by assuming that the opacity
$\sigma$ changed only on passage of the shock, and then by assuming that its subsequent
variation followed the local trend of an opacity model.
Theoretical opacities are typically tabulated over mass density $\rho$, temperature $T$,
and photon energy $E_\gamma$.
An EOS is needed to deduce $T$ along the Hugoniot and for subsequent off-Hugoniot states.
The effect of off-Hugoniot opacity variation is therefore an estimate depending on models of both
opacity and EOS, and there is no {\it a priori} guarantee of consistency between the
deduced Hugoniot and the EOS.
Ideally, the analysis would be repeated with EOS adjusted as necessary to be consistent with the
inferred Hugoniot, but such adjustment should also take account of data from other experiments,
which is beyond the scope of the work reported here.

With the opacity assumed to vary only on passage of the shock, the deduced Hugoniot locus was 
softer (lower pressure for a given compression) than when the estimated off-Hugoniot opacity variation
was included, and the opacity was deduced to drop more rapidly with pressure.
For this assessment, we primarily used {\sc sesame} EOS 7592 \cite{ses7592}
and the {\sc opal} opacity model \cite{opal}.
We also compared EOS states from a similar TF-based model,
{\sc leos} 5110 \cite{leos},
illustrating variations typical of constructions by different individuals
or using conventions and methods preferred by different research groups.
(Figs~\ref{fig:hugdp} and \ref{fig:opac}.)

\subsection{Hugoniot analysis assuming an opacity model}
\begin{figure}
\begin{center}\includegraphics[scale=0.70]{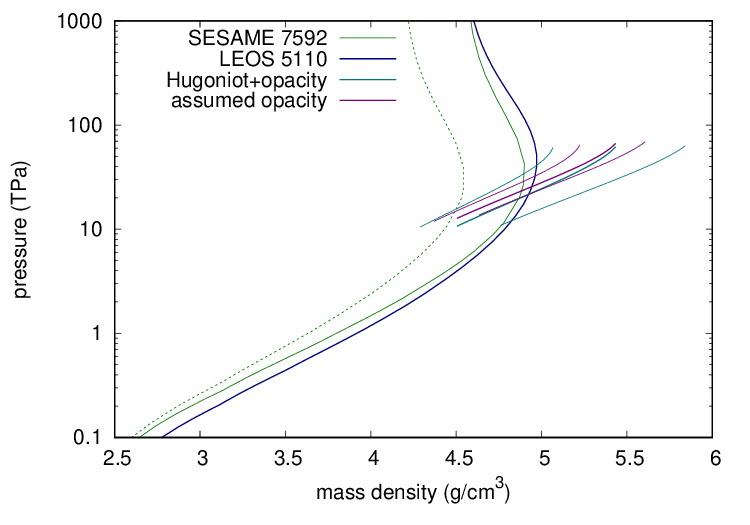}\end{center}
\caption{Hugoniot for PaMS with initial temperature 24\,K,
   deduced from the high drive shot N130103 and compared with 
   Thomas-Fermi based EOS.
   The dashed line is the {\sc sesame} Hugoniot starting at STP, for comparison.
   Thin lines are $1\sigma$ contours for the corresponding thick line.}
\label{fig:hugdp}
\end{figure}

\begin{figure}
\begin{center}\includegraphics[scale=0.70]{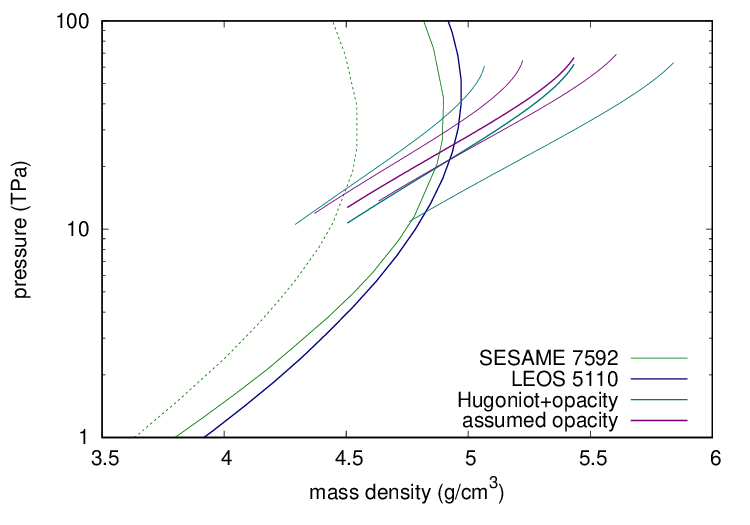}\end{center}
\caption{Hugoniot for PaMS with initial temperature 24\,K,
   deduced from the high drive shot N130103 and compared with 
   Thomas-Fermi based EOS.
   (Detail over range probed by these converging-shock experiments.)}
\label{fig:hugdpa}
\end{figure}

As with using a model to account for off-Hugoniot opacity variations, if the model is
used also for the opacity on the Hugoniot, an EOS is also needed to deduce the temperature $T$.
The deduced Hugoniot was similar to the result obtained above with opacity deduced from the radiograph
and off-Hugoniot variations also accounted for, but with significantly smaller statistical uncertainty.
(Figs~\ref{fig:hugdp} and \ref{fig:hugdpa}.)

\subsection{Opacity analysis assuming an equation of state}
\begin{figure}
\begin{center}\includegraphics[scale=0.70]{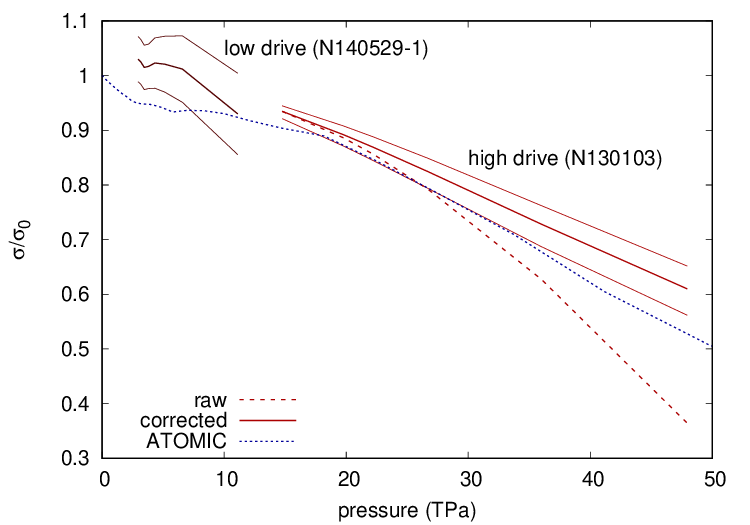}\end{center}
\caption{Opacity of PaMS to 9\,keV x-rays.
   The nominal result for shot N130103-1 assumes that the opacity changes only on passage of the shock;
   the corrected curve uses EOS and opacity models to estimate the effect of isentropic compression.
   The result for the low-drive shot N140529-1 \cite{Doeppner2018,Swift2018} is also shown,
   consistent with a constant, cold opacity.
   Thin lines are $1\sigma$ contours for the corresponding thick line.
   The dotted line is the theoretical prediction from the {\sc atomic} model
   \cite{atomic}, with Hugoniot states from a recent average-atom EOS
   {\sc leos} 5112 \cite{leos5112}.}
\label{fig:opac}
\end{figure}

Conversely to the use of the radiograph to deduce the Hugoniot for an assumed opacity, it is possible to
assume the EOS and deduce the opacity.
For inconsistencies to be avoided, the EOS must be accurate enough to reproduce the locus of the shock
given the locus of the marker layer to within the spatial resolution of the system.
(Fig.~\ref{fig:opac}.)

% ~/scidata/expts/nif/gbar/130103

% ~/scidata/expts/nif/gbar/130103

\subsection{Consistency with N130701-1}
The early disappearance of the maker layer in shot N130701-1 means that
the Hugoniot and opacity cannot be deduced simultaneously except at the lowest pressures sampled in the experiment.
The duration of the x-ray pulse was longer than in N130103-1, covering
a significantly greater range of shock radius and hence higher shock
pressures and thus greater shock heating and a decreased opacity.
Because N130701-1 was fired at ambient temperature, the initial mass density
of the PaMS was 1.085\,g/cm$^3$ rather than 1.13\,g/cm$^3$ in N130103-1,
and thus the Hugoniot pertaining to each experiment was slightly different.
The effect of changing the initial density cannot be predicted from the Hugoniot
measured at a single different density, but it can be estimated given off-Hugoniot
information such as the Gr\"uneisen parameter, which may be obtained from an existing EOS.
Similarly, the opacity at the peak pressures reached in shot N130701-1 are not
strictly constrained by the results of shot N130103-1, although the opacity
at higher pressures can be estimated by extrapolation or use of a model.

\begin{figure}
\begin{center}\includegraphics[scale=0.70]{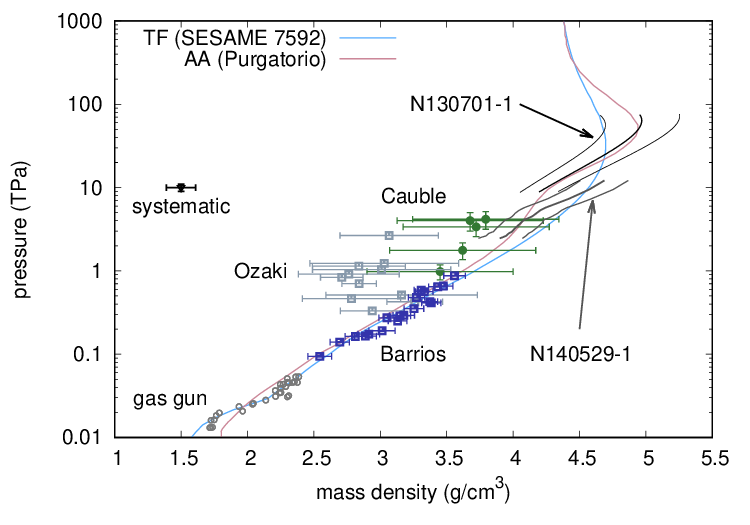}\end{center}
\caption{Ambient Hugoniot for PaMS.
   Thin lines are $1\sigma$ contours of
   statistical uncertainty from the parameterized fit to the radiograph.
   Systematic uncertainty from the uncertainty in magnification
   and the sweep speed of the x-ray streak camera is shown by the bar.
   Previous, planar, measurements
   \cite{Marsh1980,Cauble1998,Ozaki2009,Barrios2010}
   and the measurement from the low-drive spherical experiment
   N140529-1 \cite{Doeppner2018}
   are shown for comparison.}
\label{fig:hugdpamb}
\end{figure}

At early time, while the marker was visible, the radiograph was analyzed
as for N130103-1 for Hugoniot and opacity simultaneously.
The opacity was consistent with cold material.
The radiograph was therefore re-analyzed assuming the cold opacity, to give a section of the Hugoniot
with smaller uncertainty (Fig.~\ref{fig:hugdpamb}).

The radiograph was also analyzed by performing multiple one-dimensional hydrocode simulations,
assuming a model of opacity, and adjusting the EOS until the
residual in the simulated radiograph was minimized.
The opacity model was constructed from the nominal opacity deduced from
N130103-1 with off-Hugoniot variations taken from the {\sc opal} model.
The drive history was taken from radiation hydrodynamics predictions of the hohlraum temperature history,
but was allowed to vary slightly to account for inaccuracies in hohlraum energetics and ablation modeling.
Variations were introduced in the EOS in two ways: either the thermal contributions were held constant
and parameters describing the cold compression curve were varied, or the cold curve was held constant
and the thermal contributions were scaled by linear functions of mass density and temperature.
Analogously with the treatment used when accumulating Hugoniot statistics from radiographic analysis,
each simulation was assigned a likelihood from the difference in the simulated radiograph from the
measurement, and the Hugoniot locus from the corresponding EOS was used to construct a probability
distribution as a function of pressure and mass density.
The nominal best-fitting Hugoniot was extracted as the peak locus in the distribution,
along with $1\sigma$ contours for the distribution.
As with N130103-1, these contours represent the statistical uncertainty, and are correlated relatively
weakly with the systematic uncertainty from camera sweep rate and magnification.
The resulting Hugoniot did depend significantly on the opacity model used, 
but was consistent with the Hugoniot estimated directly from N130103-1 with off-Hugoniot treatment from
{\sc sesame} 7592.
The deduced Hugoniot was not sensitive to the EOS used as the basis for making variations.
(Fig.~\ref{fig:hugdpamb}.)

% ~/scidata/expts/nif/gbar/130701

\section{Conclusions}
Time-resolved x-ray radiography of a one-dimensonal shock
within a Lagrangian marker
provides a way to deduce shock Hugoniot states simultaneously with the
x-ray opacity.
The Hugoniot measurement is absolute, i.e.
without reference to an EOS standard, and can explore a range of states
in a single experiment.
For shocks driven by a laser-heated hohlraum, convergence can increase
the pressure enough that the opacity decreases significantly as
atoms in the sample are ionized.
Convergence increased the shock pressure significantly over the
externally-applied drive pressure without the drawbacks of increasing
a driving radiation intensity, such as preheat.

Experiments on PaMS explored pressures of over 70\,TPa,
though absolute Hugoniot and opacity data were inferred only up to
$\sim$40\,TPa because of the loss of the marker layer in one experiment.
The opacity inferred experimentally exhibited a decrease of similar
magnitude over a similar range of temperatures to theoretical calculations
including a detailed accounting of electron excitations. 

\section*{Acknowledgments}
Chris Mauche kindly provided tabulated {\sc opal} opacities for the x-ray energy
used here.
We acknowledge helpful discussions with Lorin Benedict, Carlos Iglesias,
John Castor, and Michael Hohensee.

This work was performed in support of
Laboratory-Directed Research and Development project 13-ERD-073
(Principal Investigator: Andrea Kritcher),
under the auspices of
the U.S. Department of Energy under contract DE-AC52-07NA27344.

\end{document}